\documentstyle[epsfig, twocolumn]{iso98}

\newcommand{\thepapername}{}

\begin{document}

\setlength{\parindent}{0pt}
\setlength{\parskip}{ 10pt plus 1pt minus 1pt}
\setlength{\hoffset}{-1.5truecm}
\setlength{\textwidth}{ 17.1truecm }
\setlength{\columnsep}{1truecm }
\setlength{\columnseprule}{0pt}
\setlength{\headheight}{12pt}
\setlength{\headsep}{20pt}
\pagestyle{veniceheadings}

\title{\bf DEEP ISOCAM OBSERVATIONS OF GALAXY CLUSTERS}

\author{{\bf D.~Fadda, D.~Elbaz} \vspace{2mm} \\
Service d'Astrophysique - CEA Saclay, F91191 Gif-sur-Yvette, France}

\maketitle

\begin{abstract}
Ten galaxy clusters, with redshift ranging from 0.2 to 1, have been
observed with the ISO camera in the two bands LW2 and LW3 (centered
respectively at 6.75~$\mu$m and 15~$\mu$m).  We present a first
analysis for three of these clusters at redshift 0.2, 0.5~and~1.
\vspace {5pt} \\


  Key~words: ISO; infrared astronomy; galaxy clusters.

\end{abstract}

\section{INTRODUCTION}
Galaxy evolution  is strongly influenced by the environment.
The galaxy morphology  is related to the density of its
environment (Dressler 1980) and,   in the  case   of galaxy  clusters,
elliptical galaxies are more common than spirals with respect to the field.
If we look to distant clusters we detect some evolution
effects as Butcher-Oemler effect (1984) or a
variation in the morphological content with the redshift (Dressler et al., 1997)
.\\

In this observational program of galaxy clusters at medium and high
redshift (P.I.: A. Franceschini; collaborators: C. Cesarsky, P.A. Duc, A. Moorwood and A. Biviano), we are interested in studying the galaxy
evolution with respect to their distance from the cluster center, in
observing active galaxies in clusters and finally, in comparing the
behavior of galaxies in clusters and in the field.  This is really
interesting especially after the results of IR deep surveys which
claim the existence of a population of galaxies with great IR fluxes
in the past ($0.6 <z < 1.$), which does not present any peculiar optical
signature (Elbaz et al. 1998, Aussel et al. 1998).

\section{OBSERVATIONS}
   Ten rich galaxy  clusters, with  redshift ranging  from 0.2 to  1,
have been observed  with the ISO camera  in the two  bands LW2 and LW3
(centered   respectively at  6.75 $\mu$m and  15  $\mu$m).
Three of them were observed very deeply (see Tables~\ref{tab:shallow} and~\ref{tab:deep}
).

Optical follow-up's of the observed clusters are scheduled in the next
months. Only that of A1689 has  already been carried out in May 1998 with
EMMI on the NTT at La Silla (P.I.: P.A. Duc). A total of about 120
spectra have been obtained  in a field of  3$\times$5 square
arcminutes  centered on Abell 1689. 
The observed objects  were mostly the optical counterpart candidates
of  the sources detected in the ISOCAM bands and the data are in
course of reduction.

\begin{table}[!h]
  \caption{\em Shallow observations}
  \label{tab:shallow}
  \begin{center}
    \leavevmode
    \footnotesize
    \begin{tabular}[h]{crrccr}
      \hline \\[-5pt]
\multicolumn{1}{c}{Cluster} & \multicolumn{1}{c}{z} & \multicolumn{1}{c}{Field
of}
& \multicolumn{1}{c}{Pix.}       & \multicolumn{1}{c}{Time}       \\ 
\multicolumn{1}{c}{Name} & \multicolumn{1}{c}{} 
& \multicolumn{1}{c}{ View}
& \multicolumn{1}{c}{FoV}       
& \multicolumn{1}{c}{(sec)}\\
\hline \\[-5pt]
A1689            &0.19 &     5.6$^\prime$ $\times$ 5.6$^\prime$  & 6$^{\prime\prime}$ & 4506  \\
GHO1600+412      &0.3  &     7.4$^\prime$ $\times$ 7.4$^\prime$  & 6$^{\prime\prime}$ & 4546\\
3C295            &0.46 &     4.0$^\prime$ $\times$ 4.0$^\prime$  & 3$^{\prime\prime}$ & 5413 \\
3C330            &0.55 &     7.4$^\prime$ $\times$ 7.4$^\prime$  & 6$^{\prime\prime}$ & 4536 \\
CL0016+1609      &0.55 &     5.6$^\prime$ $\times$ 5.6$^\prime$  & 6$^{\prime\prime}$ & 4476 \\
J1888.16CL       &0.56 &     5.6$^\prime$ $\times$ 5.6$^\prime$  & 6$^{\prime\prime}$ & 4506 \\
CL1322+3029      &0.7  &     2.2$^\prime$ $\times$ 2.2$^\prime$  & 3$^{\prime\prime}$ & 5373 \\
GHO1322+3027     &0.75 &     4.0$^\prime$ $\times$ 4.0$^\prime$  & 6$^{\prime\prime}$ & 2732\\
GHO1603+4313     &0.895&     2.2$^\prime$ $\times$ 2.2$^\prime$  & 3$^{\prime\prime}$ & 5373\\
CL1603+4329      &0.92 &     4.0$^\prime$ $\times$ 4.0$^\prime$  & 6$^{\prime\prime}$ & 2732\\
CL1415+5244      &...  &      Failed      & ... & ... \\
      \hline \\
      \end{tabular}
  \end{center}
\end{table}

\begin{table}[!h]
  \caption{\em Deep observations}
  \label{tab:deep}
  \begin{center}
    \leavevmode
    \footnotesize
    \begin{tabular}[h]{crcccr}
      \hline \\[-5pt]
\multicolumn{1}{c}{Cluster} & \multicolumn{1}{c}{z} & \multicolumn{1}{c}{Band}
& \multicolumn{1}{c}{Field of}  &   \multicolumn{1}{c}{Pix.}   & \multicolumn{1}{c}{Time}       \\ 
\multicolumn{1}{c}{Name} & \multicolumn{1}{c}{} &\multicolumn{1}{c}{} 
& \multicolumn{1}{c}{View}       
& \multicolumn{1}{c}{FoV}
& \multicolumn{1}{c}{(sec)}       \\ 
\hline \\[-5pt]
3C330        & 0.55 & LW3 &15.2$^\prime \times$3.3$^\prime$ & 6$^{\prime\prime}$ &  20584\\
J1888.16CL   & 0.56 & LW2 &13.7$^\prime \times$3.35$^\prime$& 6$^{\prime\prime}$ &  20967\\
J1888.16CL   & 0.56 & LW3 &15.2$^\prime \times$3.3$^\prime$ & 6$^{\prime\prime}$ &  20624\\
GHO1603+4313 & 0.895& LW2 &13.7$^\prime \times$3.35$^\prime$& 6$^{\prime\prime}$ &  20967\\
GHO1603+4313 & 0.895& LW3 &15.2$^\prime \times$3.3$^\prime$ & 6$^{\prime\prime}$ &  20604\\
      \hline \\
      \end{tabular}
  \end{center}
\end{table}

\section{IR EMISSION IN ISOCAM FILTERS}
The part of the galaxy spectrum seen by each ISOCAM filter
is a function of the redshift (K correction).

The galaxy spectrum is the sum of three different components: UIBs, warm
dust and forbidden lines of ionized gas.
In the case of nearby clusters, the LW3 band is centered on the
warm dust emission  and the LW2 band is dominated by the contribution
of the UIBs.
\begin{figure*}[!ht]
  \begin{center}
    \leavevmode
   \centerline{\hspace{5.5cm}\epsfig{file=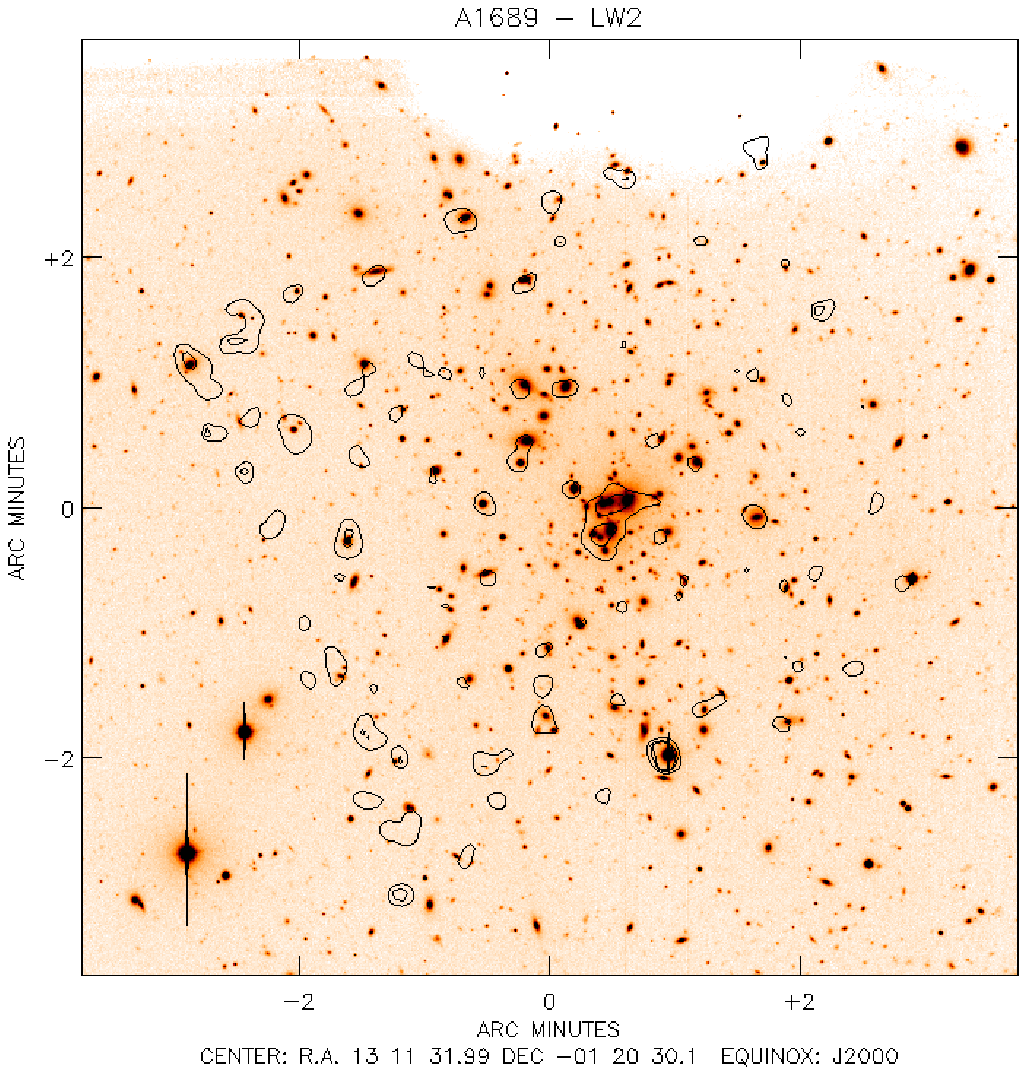,width=14.0cm}\hspace{-5.5cm}
\epsfig{file=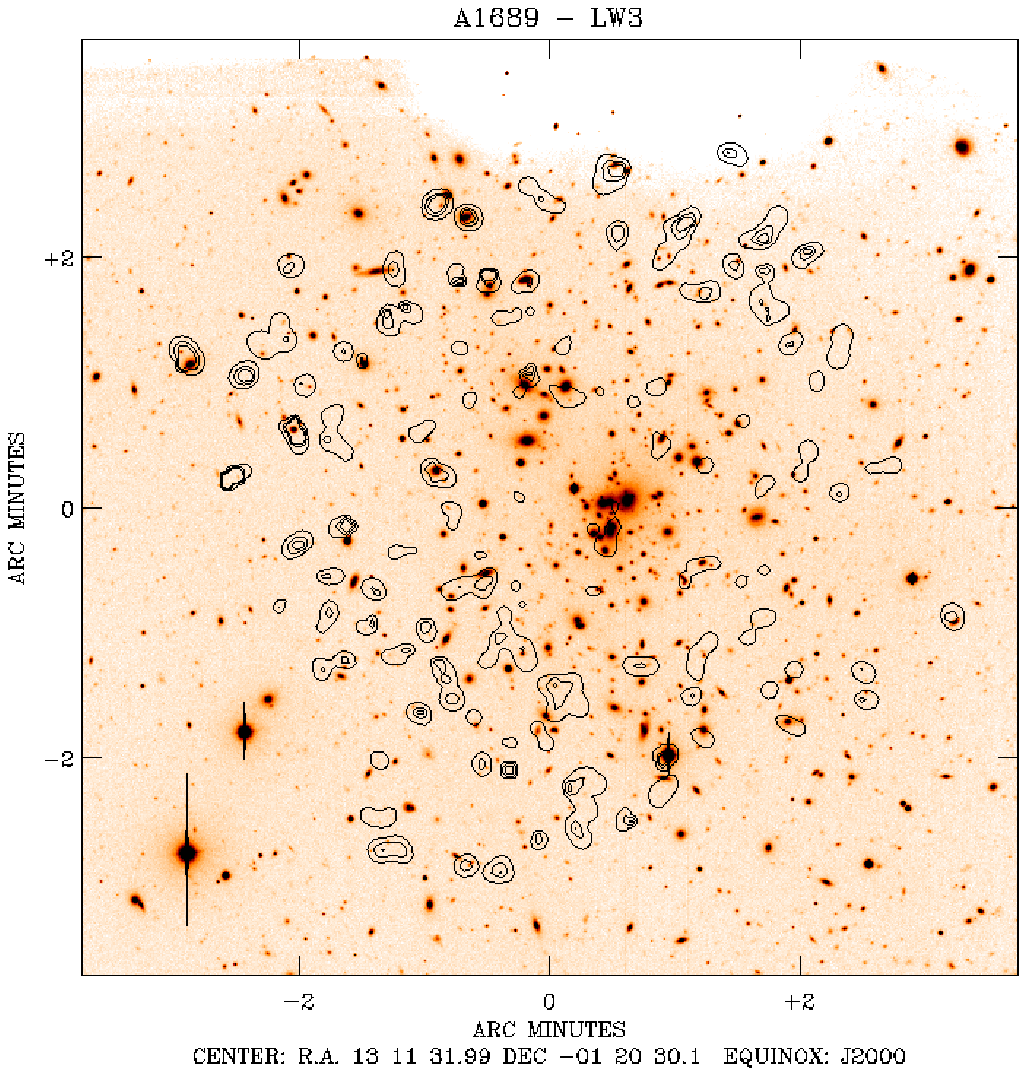,clip=,width=14.0cm}
}
  \end{center}
  \caption{\em A1689 in the LW2 (left) and LW3 (right) bands superposed on a 
NTT image (PI: P.A. Duc)}
  \label{fig:a1689-images}
\end{figure*}

\begin{figure}[!h]
   \begin{center}
   \leavevmode
   \centerline{\epsfig{file=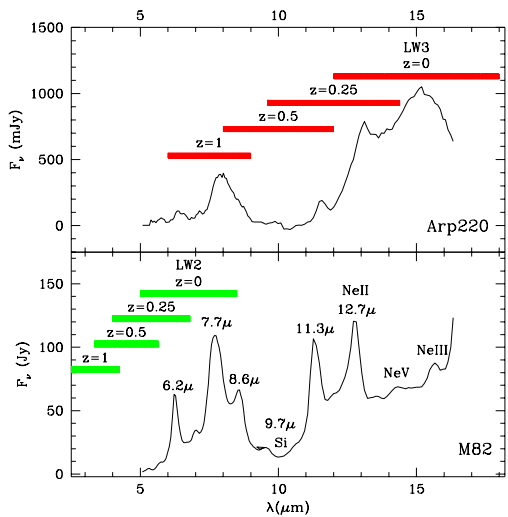,
               width=7.0cm}}
   \end{center}
 \caption{\em The figure shows the spectra of two known starburst galaxies
observed  by ISO (courtesy of D. Tran  and V. Charmandaris) and the LW2 and LW3 bands at different redshifts.}
 \label{fig:lw2lw3}
 \end{figure}

When we consider distant clusters ($0.5 < z < 1.0$), the LW3 band is
more  and more contaminated by the UIBs, while in the LW2 band the
contribution of the stellar continuum, especially from old population
stars, overtakes the UIB features (see Fig.~\ref{fig:lw2lw3}).

\section{DISCUSSION}
\label{sec:commands}
We detect a spatial segregation in the distribution of ISOCAM sources.
\begin{figure}[!h]
   \begin{center}
   \leavevmode
\vspace{-1.5cm}
\epsfig{file=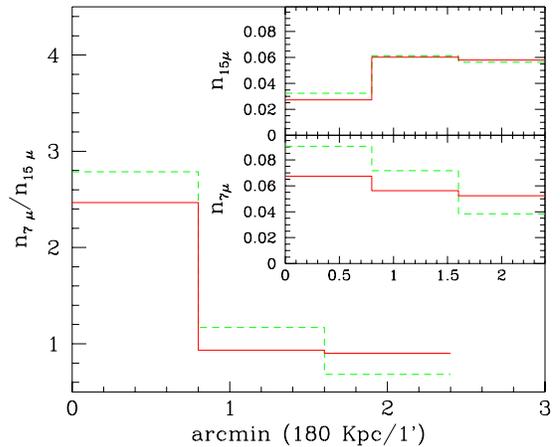,
               width=7.0cm,angle=-90}
   \end{center}
 \caption{\em Source number density for A1689. Shaded and solid lines refer respectively to 5$\tau_W$ and 7$\tau_W$ thresholds in the wavelet detections of sources (Starck et al., 1998).}
 \label{fig:a1689-dens}
 \end{figure}
In the A1689 field (Fig.~\ref{fig:a1689-images})
we can remark an overdensity of LW2 sources in the cluster center,
while LW3 sources seem to avoid the central region.
There are more  detections associated to the dust (LW3  band) in  the
outskirts than in the central region, likely because galaxies which
are far from the cluster center are generally more active than central
galaxies (see Fig.~\ref{fig:a1689-dens}).
\begin{figure}[!t]
   \begin{center}
   \leavevmode
\epsfig{file=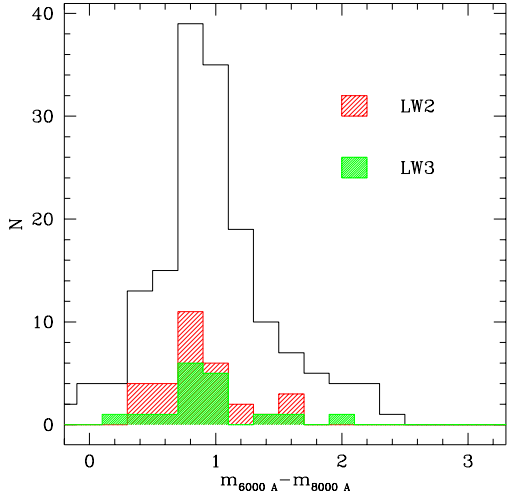,
               width=7.0cm}
   \end{center}
 \caption{\em Color distribution for A1689.}
 \label{fig:a1689-colors}
 \end{figure}

If we compare the distribution of colors for the galaxies (Gudehus \&
Hegyi 1991) with the analogous distributions for galaxies detected in
the two bands (Fig.~\ref{fig:a1689-colors}), we note that
these distributions do  not differ significatively, as in the case of 
ISOCAM observations of the Hubble Deep Field (Aussel et al. 1998). This shows
that the activity unveiled by the mid-infrared images is hidden in the optical,
corroborating the observation of the Antennae galaxy by Mirabel et al.~(1998).
\begin{figure}[!h]
   \begin{center}
   \leavevmode
\vspace{-1.5cm}
   \centerline{\epsfig{file=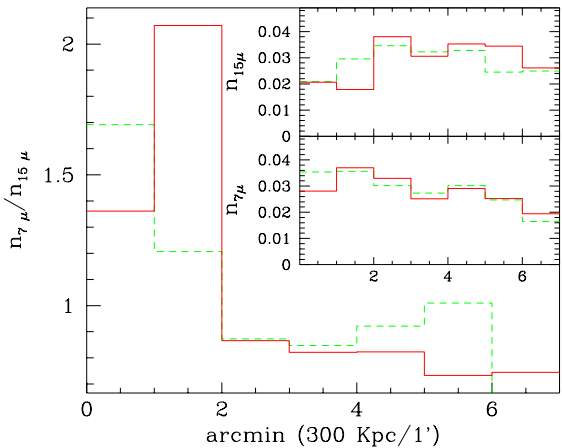,
               width=7.0cm,angle=-90}}
   \end{center}
 \caption{\em  Source number density for J1888.16CL. Shaded and solid lines refer respectively to 5$\tau_W$ and 7$\tau_W$ thresholds in the wavelet detections of sources (Starck et al., 1998).}
 \label{fig:J1888-dens}
 \end{figure}

\begin{figure}[!t]
   \begin{center}
   \leavevmode
\vspace{-0.7cm}
   \centerline{\epsfig{file=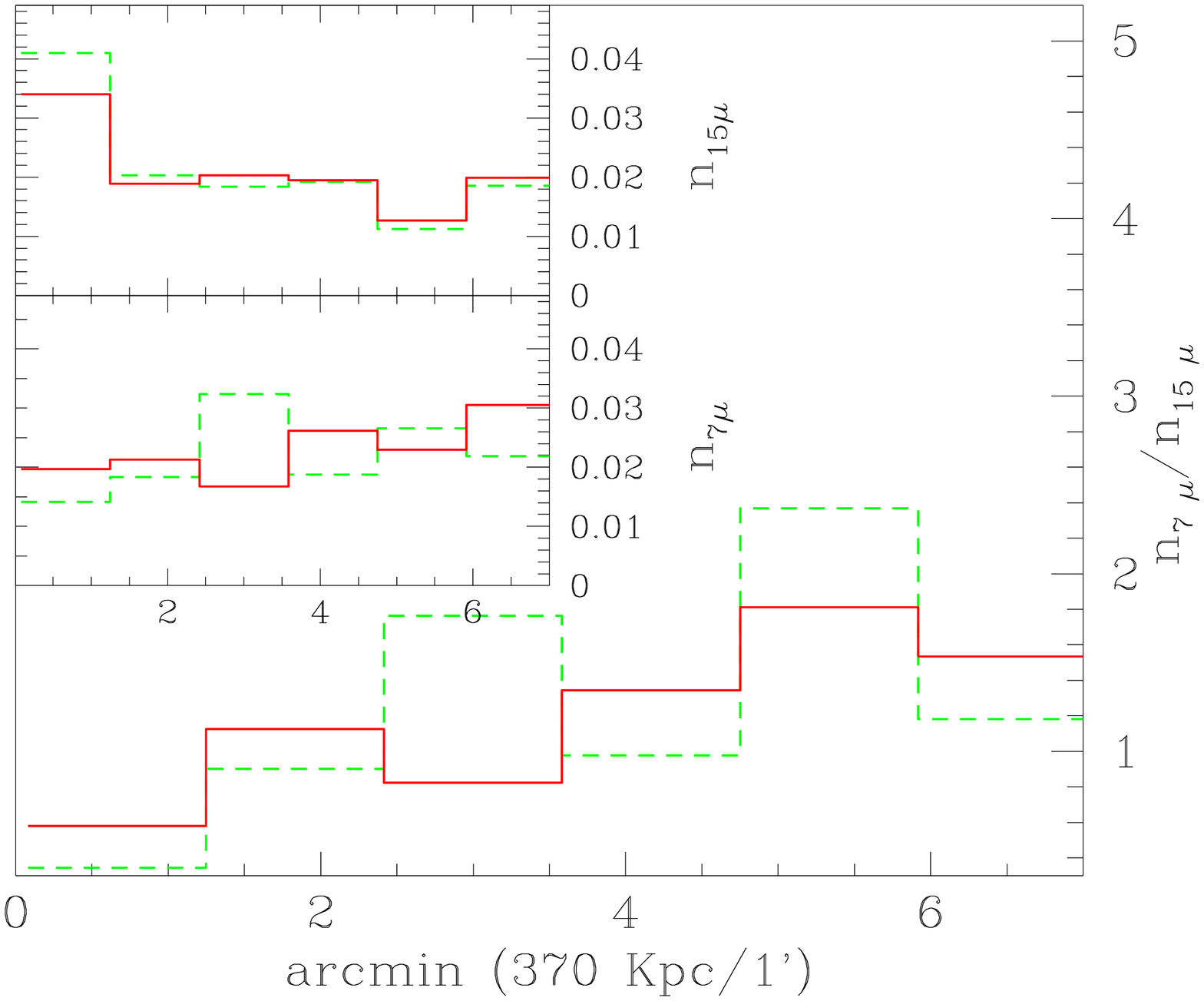,
               width=7.0cm,angle=-90}}
   \end{center}
 \caption{\em  Source number density for GHO 1603+4313. Shaded and solid lines refer respectively to 5$\tau_W$ and 7$\tau_W$ thresholds in the wavelet detections of sources (Starck et al., 1998).} 
 \label{fig:gho-dens}
 \end{figure}

The segregation effect is more clearly visible in the cluster
J1888.16CL at a redshidft of 0.5 (Fig.~\ref{fig:J1888-dens}). This effect could be due to a more
intense activity of galaxies or to a favored detection of arclets at
15 $\mu$m as in the case of A2390 (Altieri \& al, 98). 
For the cluster GHO 1603+4313 at $z=1$ we find an uniform distribution for 
the LW2 band and an overdensity of galaxies at the cluster center for the LW3 band (Fig.~\ref{fig:gho-dens}). This is consistent with previous cases, since LW3 at $z=1$ corresponds to LW2 band at rest ($z=0$), due to k-correction, but may be also associated to a stronger star formation activity at higher $z$.

\section*{ACKNOWLEDGMENTS}
We thank H. Aussel for his helpful advises.\\
The ISOCAM data presented in this paper was analyzed using "CIA", 
a joint development by the ESA Astrophysics Division and the ISOCAM
Consortium led by the ISOCAM PI, C. Cesarsky, Direction des Sciences de la
Matiere, C.E.A., France.

\end{document}